\documentclass[11pt]{article}
\usepackage{moriond,epsfig}

\def\be{\begin{equation}}
\def\ee{\end{equation}}
\def\bea{\begin{eqnarray}}
\def\eea{\end{eqnarray}}

\begin{document}
\vspace*{4cm}
\title{Diffractive production of $c \bar c$}

\author{Marta {\L}uszczak $^{2}$, Rafa{\l} Maciu{\l}a $^{1}$ , Antoni Szczurek $^{1,2}$}

\address{$^{1}$ {\em Institute of Nuclear Physics PAN\\
PL-31-342 Cracow, Poland\\}
$^{2}$ {\em University of Rzesz\'ow\\
PL-35-959 Rzesz\'ow, Poland\\}
}

\maketitle\abstracts{
At high-energies the gluon-gluon fusion is the dominant mechanism of 
$c \bar c$ production. This process was calculated in the NLO collinear 
as well as in the k$_t$-factorization approaches in the past.
In this presentation we concentrate on production of $c \bar c$ pairs including several
subleading mechanisms. 
This includes:
$gg \to Q \bar Q$, $\gamma g \to Q \bar Q$, $g \gamma \to Q \bar Q$, 
$\gamma \gamma \to Q \bar Q$.
In this context we use MRST-QED 
parton distributions which include photon as a parton in the proton
as well as elastic photon distributions calculated in the equivalent photon approximation.
We present distributions in the $c$ quark ($\bar c$ antiquark) rapidity 
and transverse momenta and compare them to the dominant gluon-gluon fusion
contribution.valent photon approximation.
We discuss also inclusive single and central diffractive processes using
diffractive parton distribution found from the analysis of HERA diffractive data.
As in the previous case we present distribution in $c$ ($\bar c$)
rapidity and transverse momentum.
Next we present results for exclusive central diffractive mechanism
discussed recently in the literature. We show corresponding differential
distributions and compare them with corresponding distributions for
single and central diffractive components.
Finally we discuss production of two pairs of $c \bar c$ within a simple
formalism of double-parton scattering (DPS). Surprisingly very large cross sections,
comparable to single-parton scattering (SPS) contribution, are predicted for LHC
energies.
}
\section{Introduction}

In this presentation we discuss contributions of some
subleading mechanisms neglected in the analysis of $c \bar c$
production. We include contributions of photon-gluon
(gluon-photon) as well as purely electromagnetic
photon-photon fusion. Here we present only some selective results.
The formalism and more details has been shown and discussed
elsewhere \cite{LMS2011}.

We discuss also diffractive processes (single and central)
in the framework of Ingelman-Schlein model corrected for absorption.
Such a model was used in the estimation of several diffractive
processes \cite{diffractive_dijets,double_pomeron,double_diff,H1}.

The absorption corrections are necessary to understand a huge 
Regge-factorization breaking observed in single and central 
production at Tevatron.

\section{Production of heavy quarks}

The cross section for the $c \bar c$ production, assuming gluon-gluon
fusion, was calculated both in collinear and $k_t$ factorization
approaches. Our group has done detailed calculations in the second approach
(see e.g. \cite{luszczak1,luszczak2}).

In the leading-order (LO) approximation within the $k_t$-factorization approach
the quadruply differential cross section in the rapidity 
of $Q$ ($y_1$), in the rapidity of $\bar Q$ ($y_2$) and in the 
transverse momentum of $Q$ ($p_{1,t}$) and $Q$ ($p_{2,t}$) can 
be written as \cite{luszczak1,luszczak2}
\scriptsize{
\begin{eqnarray}
\nonumber 
\frac{d \sigma}{d y_1 d y_2 d^2p_{1,t} d^2p_{2,t}} =
\sum_{i,j} \; \int \frac{d^2 \kappa_{1,t}}{\pi} \frac{d^2 \kappa_{2,t}}{\pi}
\frac{1}{16 \pi^2 (x_1 x_2 s)^2} \; 
\\
\nonumber 
\overline{ | {\cal M}_{ij \to Q \bar Q} |^2}
\delta^{2} \left( \vec{\kappa}_{1,t} + \vec{\kappa}_{2,t} 
                 - \vec{p}_{1,t} - \vec{p}_{2,t} \right) \;
\\
\nonumber 
{\cal F}_i(x_1,\kappa_{1,t}^2) \; {\cal F}_j(x_2,\kappa_{2,t}^2) \; , 
\end{eqnarray}
}
\normalsize{
where ${\cal F}_i(x_1,\kappa_{1,t}^2)$ and ${\cal F}_j(x_2,\kappa_{2,t}^2)$
are the so-called unintegrated gluon (parton) distributions. 
The unintegrated parton distributions must be evaluated at:
\begin{eqnarray}
x_1 &=& \frac{m_{1,t}}{\sqrt{s}}\exp( y_1) 
      + \frac{m_{2,t}}{\sqrt{s}}\exp( y_2),
\nonumber \\
x_2 &=& \frac{m_{1,t}}{\sqrt{s}}\exp(-y_1) 
      + \frac{m_{2,t}}{\sqrt{s}}\exp(-y_2),
\nonumber
\end{eqnarray}
where $m_{i,t} = \sqrt{p_{i,t}^2 + m_Q^2}$.
}

\section{Photon induced production of heavy quarks}

The dominant contributions of heavy quark-antiquark production are initiated 
by gluon-gluon fusion or quark-antiquark annihilation. In general,
even photon can be a constituent of the proton. This idea was considered 
by Martin, Roberts, Stirling and Thorne in Ref.\cite{MRST04}.

\begin{figure}[!htp]
\begin{center}
\includegraphics[width=1.7cm]{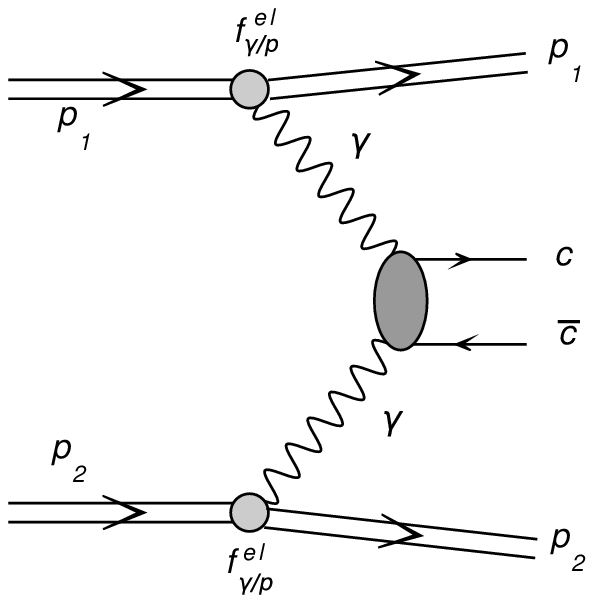}
\includegraphics[width=1.7cm]{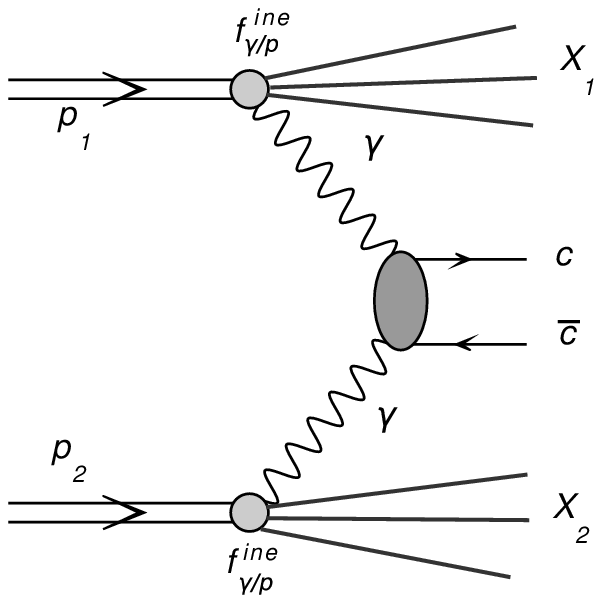}
\includegraphics[width=1.7cm]{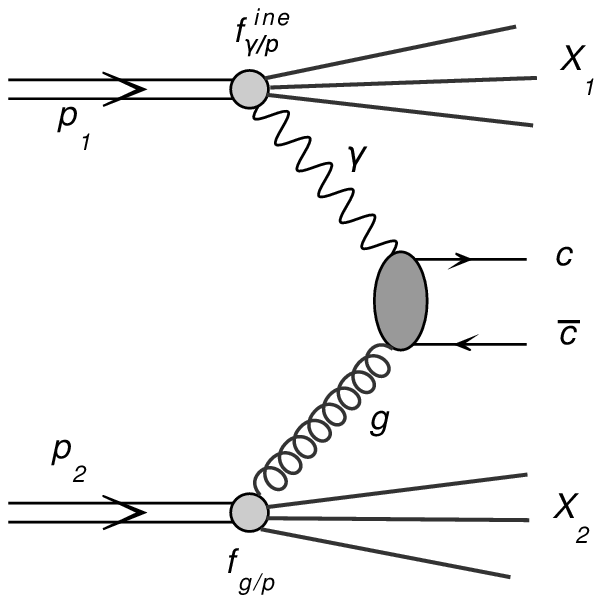}
\includegraphics[width=1.7cm]{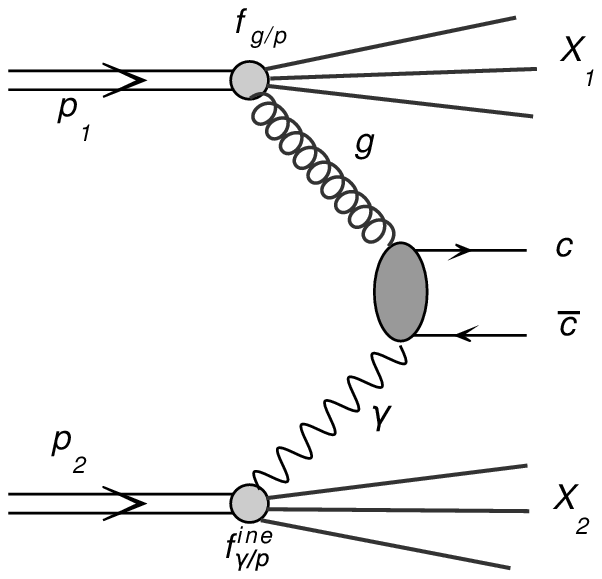}\\
\includegraphics[width=1.7cm]{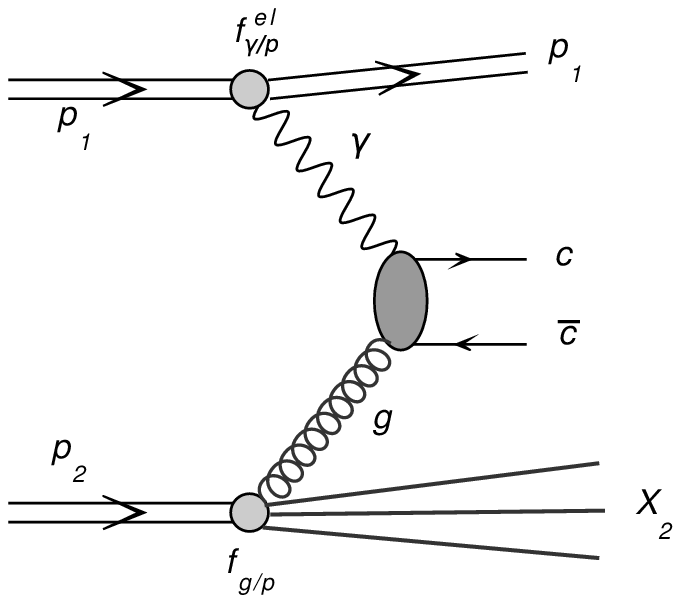}
\includegraphics[width=1.7cm]{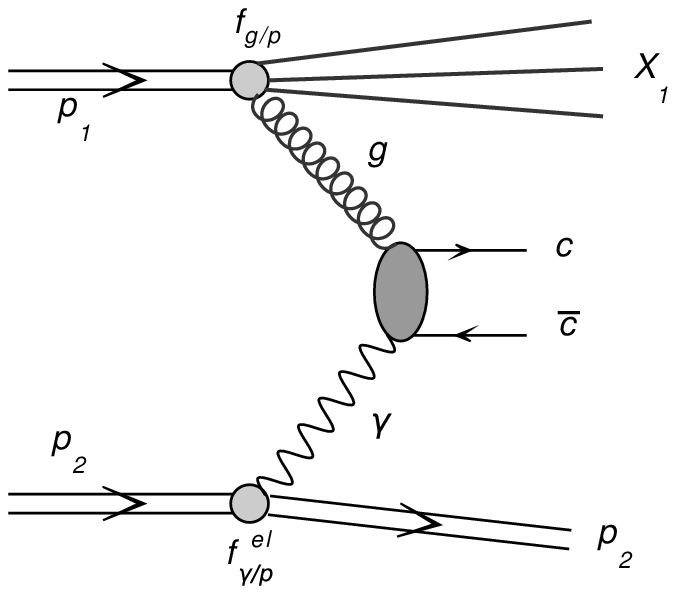}
\includegraphics[width=1.7cm]{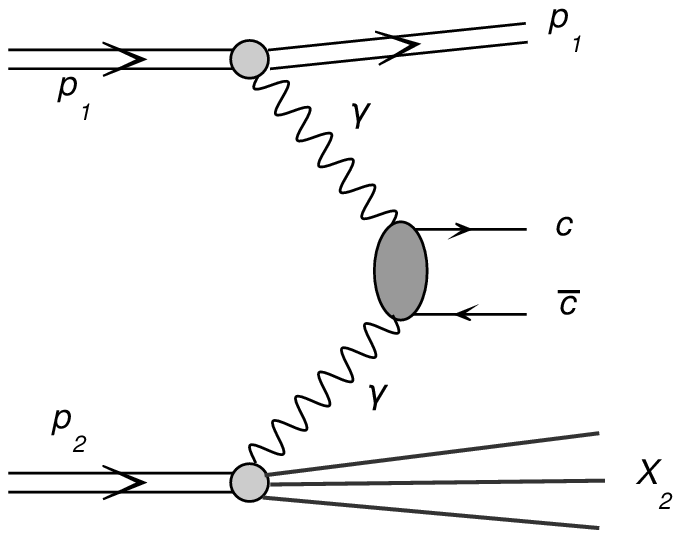}
\includegraphics[width=1.7cm]{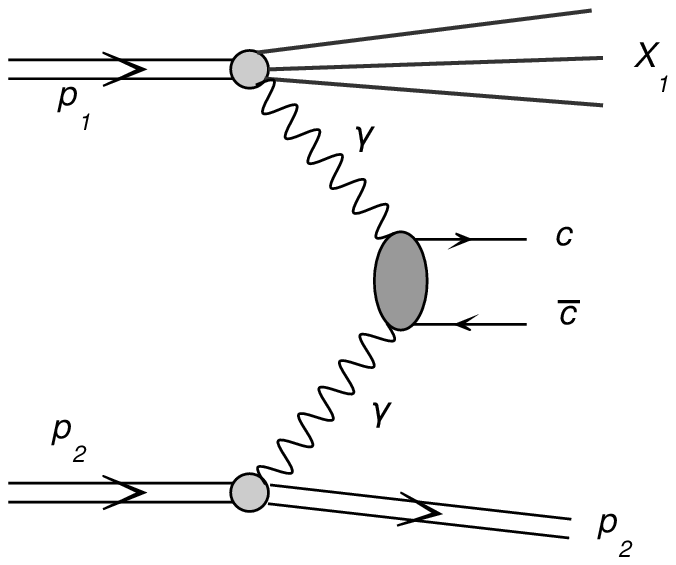}
\caption{Diagrams representing photon-induced mechanisms 
of heavy quark production.
}
\label{fig:new_diagrams}
\end{center}
\end{figure}

If the photon is a constituent of the nucleon then other mechanisms
of $c \bar c$ production presented in Fig.\ref{fig:new_diagrams} 
are possible.

\section{Results}

\subsection{Gluon-gluon fusion}

\begin{figure} [!thb]
\begin{center}
\includegraphics[width=3.7cm]{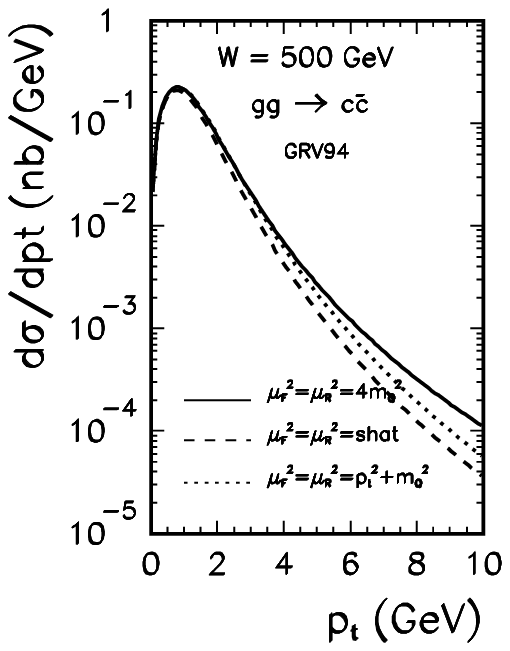}
\includegraphics[width=3.7cm]{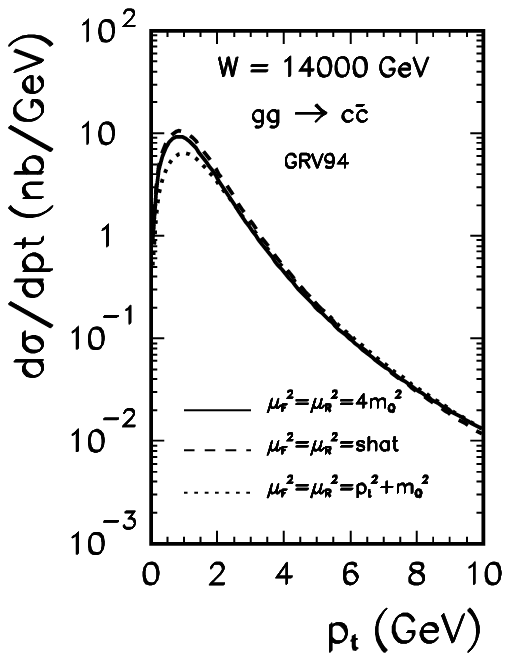}
\caption{Distribution in quark/antiquark 
transverse momentum at $\sqrt{s}$ = 500 GeV (left panel) and for 
$\sqrt{s}$ = 14 TeV (right panel) for different choices of 
scales and for GRV gluon distribution.
}
\label{fig:dsig_dpt_gg}
\end{center}
\end{figure}

Before we go to the new mechanisms we will present results for the dominant
gluon-gluon fusion.
In Fig.\ref{fig:dsig_dpt_gg} we show distributions in transverse 
momentum of $c$ (or $\bar c$) for the gluon-gluon fusion mechanism
for different popular choices of scales ($\mu^2 = 4 m_c^2, M_{c \bar c}^2,
p_t^2 + m_c^2$). We show our results for $\sqrt{s}$ = 500 GeV (left panel)
and $\sqrt{s}$ = 14 TeV (right panel). In this calculation we have used
GRV \cite{GRV94} PDFs. The figure shows typical uncertainties due to the
choice of the scale. We wish to stress here that at the higher energies
the results of the calculations depend on the gluon distributions at
small values of $x$. 

\subsection{$\gamma g$ and $g \gamma$ subprocesses}

\begin{figure} [!thb]
\begin{center}
\includegraphics[width=3.5cm]{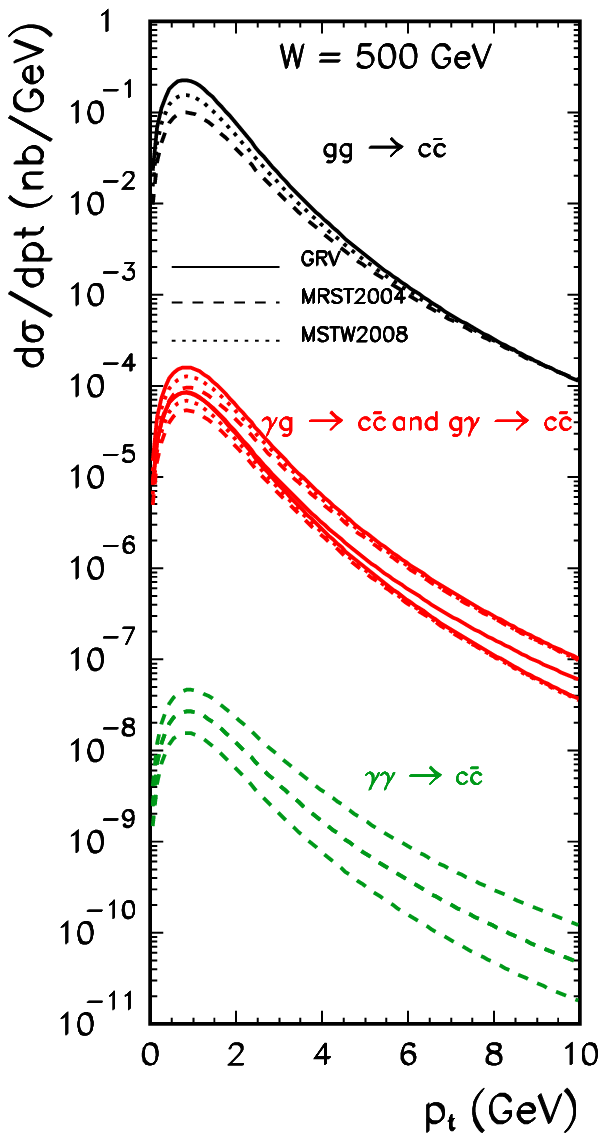}
\includegraphics[width=3.5cm]{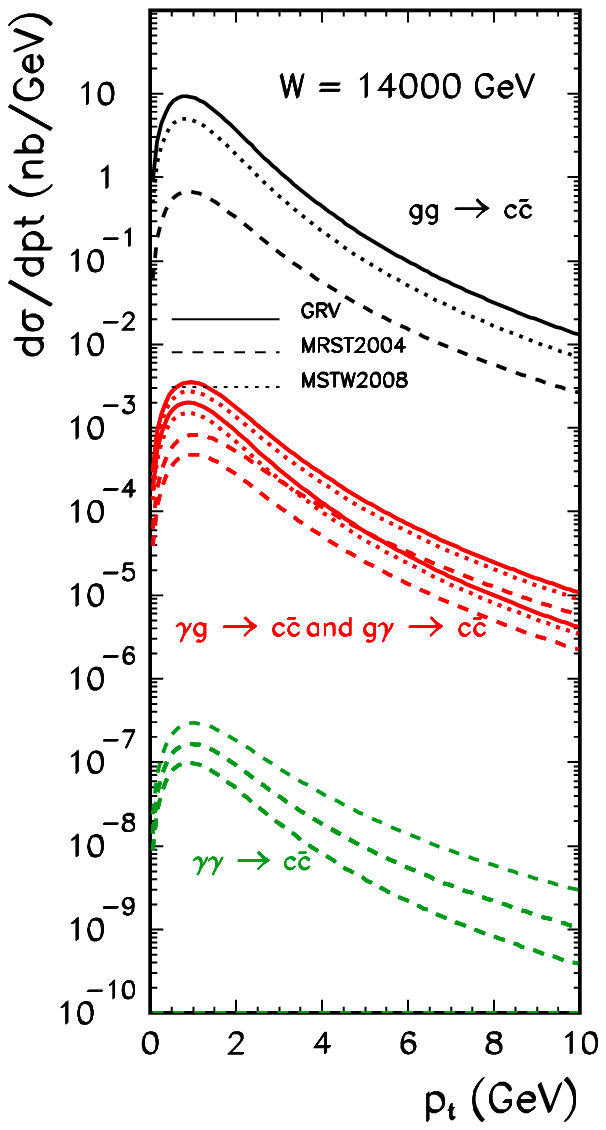}
\caption{
Transverse momentum distribution for the standard gluon-gluon
mixed gluon-photon and photon-gluon as well as for photon-photon
contributions for RHIC (left) and LHC (right).
}
\label{fig:dsig_dpt_gamma}
\end{center}
\end{figure}

In Fig.\ref{fig:dsig_dpt_gamma}
we show transverse momentum distributions for the dominant gluon-gluon 
as well as for the subleading photon-gluon (gluon-photon) and
photon-photon components for different gluon distribution functions 
\cite{GRV94,MRST04,MSTW08}  for
the RHIC energy $\sqrt{s}$ = 500 GeV and for the nominal LHC energy 
$\sqrt{s}$ = 14 TeV, respectively.
At the LHC energy the results for different GPDs differ considerably which
is a consequence of the poorly known small-x region.
The differences at the energy $\sqrt{s}$ = 14 TeV are 
particularly large which can be explained by the fact that a product of
gluon distributions (both at small x) enters the cross section formula.
New measurement of $c \bar c$ at the nominal LHC energy will be
therefore a severe test of gluon distributions at small $x$ and not too
high factorization scales not tested so far. Similar uncertainties for 
the $\gamma g$ and $g \gamma$ are smaller as here only one gluon 
distribution appears in the corresponding cross section formula.
The uncertainties for the photon distributions are not yet quantified.

It is very difficult to quantify uncertainties related to photon PDFs
as only one set of PDFs includes photon as a parton of the proton.
Here the isospin symmetry violation (not well known at present)
would be an useful limitation.
Our collection of the results for the photon induced mechanisms show
that they are rather small and their identification would be rather
difficult as the different distributions are very similar to those
for the gluon-gluon fusion.
Our intension is to document all the subleading terms.
Our etimation shows that the sum of all the photon induced terms
is less than 0.5 \% and is by almost 2 orders of magnitude smaller
than the uncertainties of the dominant leading-order
gluon-gluon term.

\section{Single and central diffraction}

\subsection{Formalism}

\begin{figure} [!thb]
\begin{center}
\includegraphics[width=3cm]{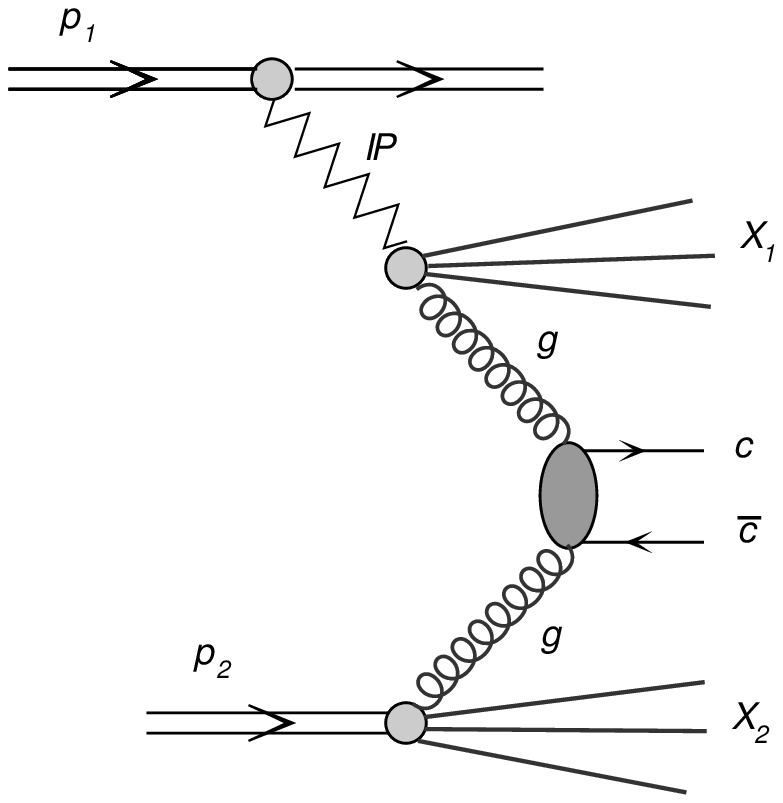}
\includegraphics[width=3cm]{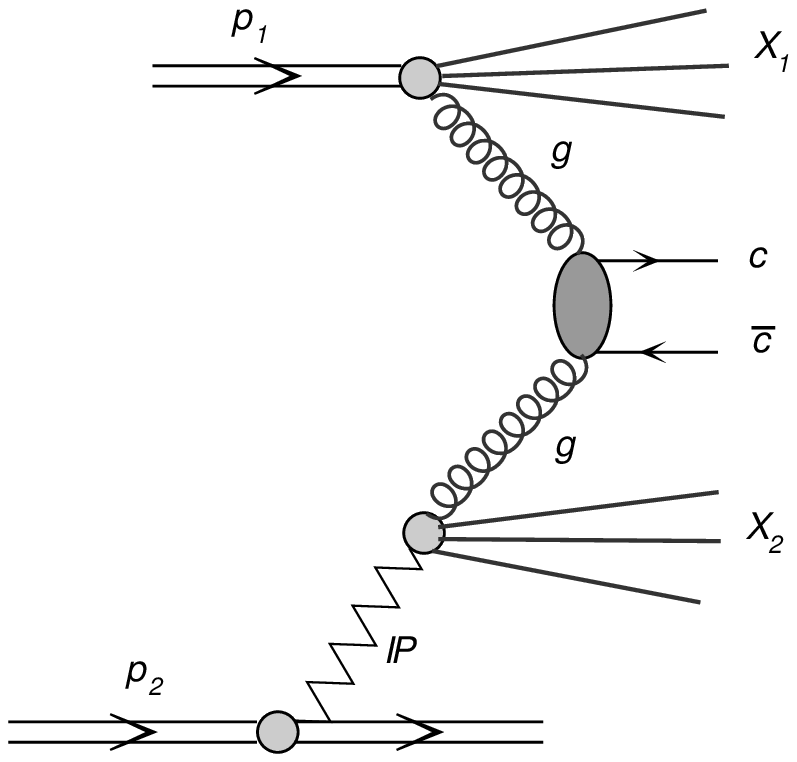}
\caption{
The mechanism of single-diffractive production of $c \bar c$.
}
\label{fig:ccbar_sd}
\end{center}
\end{figure}

\begin{figure} [!thb]
\begin{center}
\includegraphics[width=3cm]{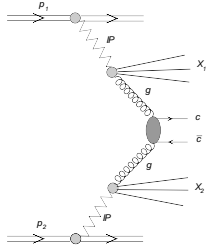}
\caption{The mechanism of central-diffractive production of $c \bar c$.
}
\label{fig:ccbar_dd}
\end{center}
\end{figure}

The mechanisms of the diffractive production 
of heavy quarks ($c \bar c$) are shown in 
Figs.\ref{fig:ccbar_sd}, \ref{fig:ccbar_dd}.
The formalism how to calculate respective cross section has been presented elsewhere \cite{LMS2011}.
 
\subsection{Results}

\begin{figure} [!thb]
\begin{center}
\includegraphics[width=4cm]{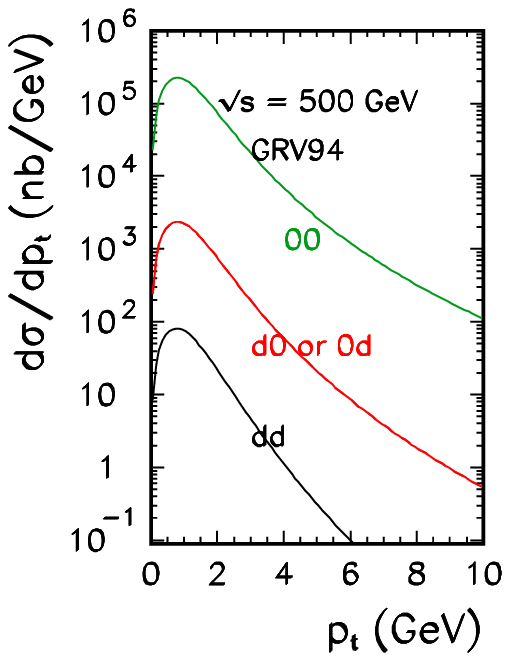}
\includegraphics[width=4cm]{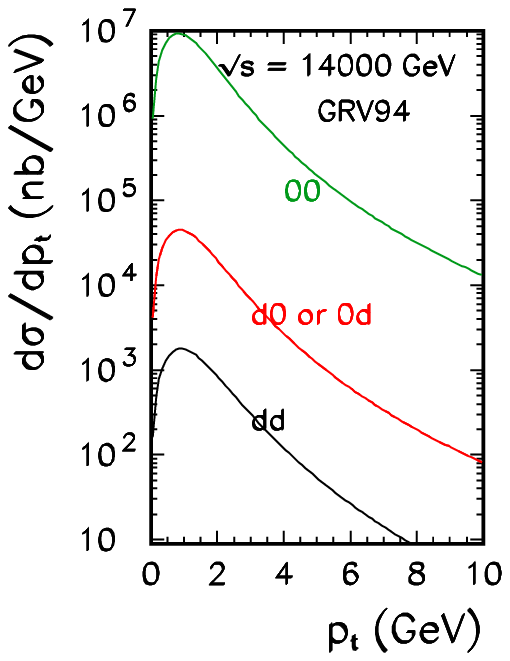}
\caption{Transverse momentum distribution of $c$ quarks (antiquarks)
for RHIC energy $\sqrt{s} =$ 500 GeV (left panel) and for LHC energy
$\sqrt{s}$ = 14 TeV (right panel) for the GRV94 gluon distributions.
The result for single diffractive (0d or d0), central diffractive (dd) 
mechanisms are compared with that for the standard gluon-gluon fusion (00).
}
\label{fig:diff_dsig_dpt_diff}
\end{center}
\end{figure}

In Fig.\ref{fig:diff_dsig_dpt_diff} we show transverse momentum
distributions of charm quarks (or antiquarks). The distribution
for single diffractive component is smaller than that for the
inclusive gluon-gluon fusion by almost 2 orders of magnitude. Our
results include gap survival factor \cite{LMS2011}. 
The cross section for central diffractive component is smaller by 
additional order of magnitude.


\begin{figure} [!thb]
\begin{center}
\includegraphics[width=4cm]{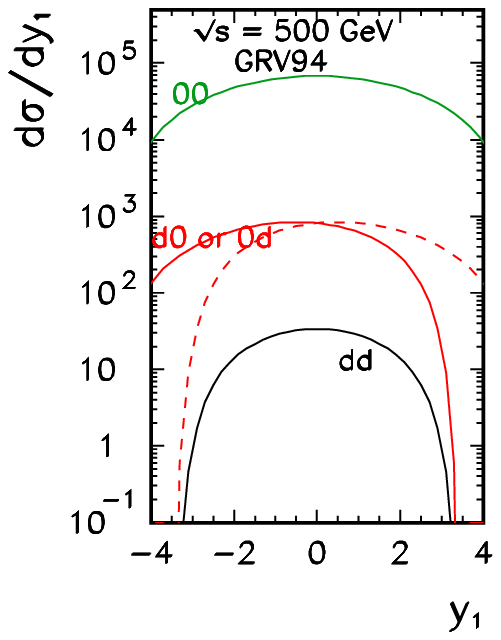}
\includegraphics[width=4cm]{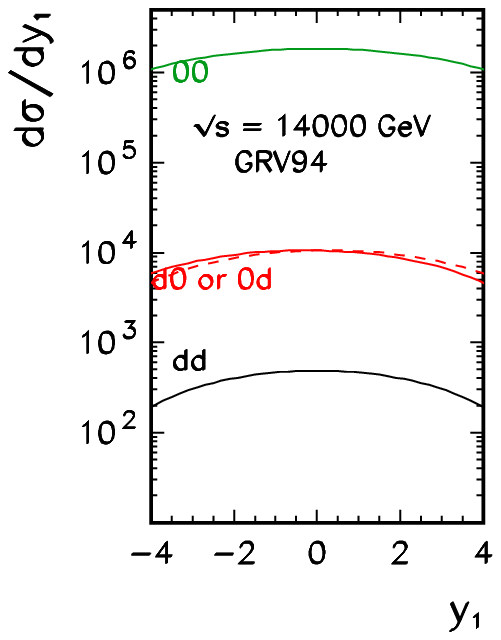}
\caption{
Rapidity distribution of $c$ quarks (antiquarks)
for RHIC energy $\sqrt{s} =$ 500 GeV (left panel) and 
for LHC energy $\sqrt{s}$ = 14 TeV (right panel) for the GRV94 gluon 
distributions. The result for single diffractive (0d or d0), 
central diffractive (dd) mechanisms
are compared with that for the standard gluon-gluon fusion (00).
}
\label{fig:diff_dsig_dy1_diff}
\end{center}
\end{figure}

In Fig.\ref{fig:diff_dsig_dy1_diff} we show distributions in quark
(antiquark) rapidity. We show separately contributions of two different 
single-diffractive components (which give the same distributions in
transverse momentum) and the contribution of central-diffractive
component in Fig.\ref{fig:diff_dsig_dpt_diff}.
When added together the single-diffractive components produce a 
distribution in rapidity similar in shape to that for the standard inclusive case.


The cross section for single and central diffraction is rather small
compared to the dominant gluon-gluon fusion component.
However, a very specific final state should allow its identification
by imposing special conditions on the one-side (single-diffractive
process) and both-side (central diffractive process) rapidity gaps.
We hope that such an analysis is possible at LHC. Special care
should be devoted to the observation of the exclusive $c \bar c$
production.
Without a special analysis of the final state multiplicity
the exclusive $c \bar c$ production may look like an inclusive
central diffraction.

\section{Production of two $c \bar c$ pairs in double-parton scattering}

The general formula for the cross section 
in terms of double-parton distributions can be written \cite{LMS2012}:
\begin{eqnarray}
d \sigma^{DPS} &=& \frac{1}{2 \sigma_{eff}}
F_{gg}(x_1,x_2,\mu_1^2,\mu_2^2) F_{gg}(x'_{1}x'_{2},\mu_1^2,\mu_2^2)
\nonumber \\
&&d \sigma_{gg \to c \bar c}(x_1,x'_{1},\mu_1^2)
d \sigma_{gg \to c \bar c}(x_2,x'_{2},\mu_2^2) \; dx_1 dx_2 dx'_1 dx'_2 \, .
\label{cs_via_doublePDFs}
\end{eqnarray}

\begin{figure} [!thb]
\begin{center}
\includegraphics[width=6.5cm]{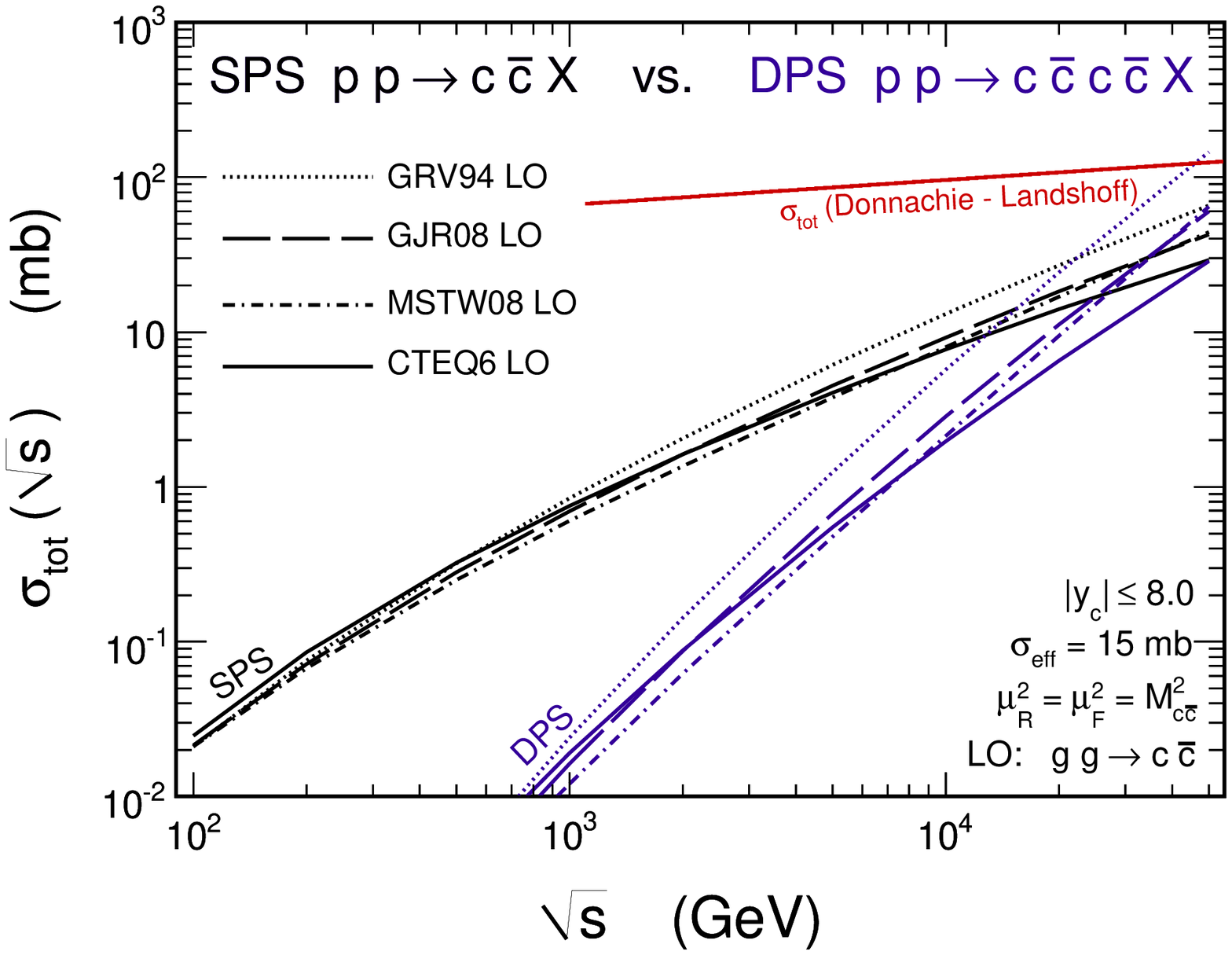}
\includegraphics[width=6.5cm]{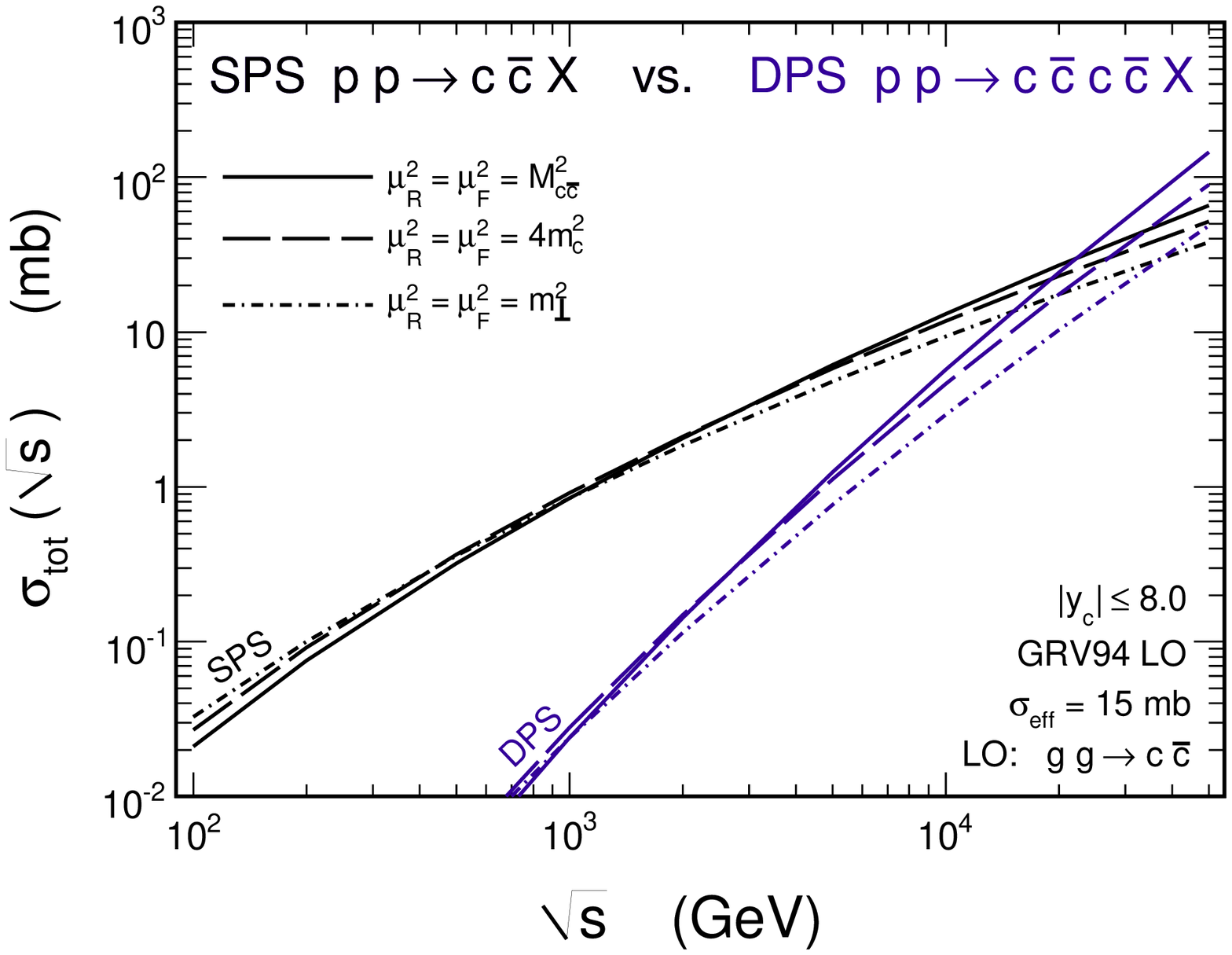}
\caption{
Total LO cross section for single-parton and double-parton
scattering as a function of center-of-mass energy (left panel) and 
uncertainties due to the choice of (factorization, renormalization) scales (right panel). 
We show in addition a parametrization of the total cross section in the left panel. 
}
\label{fig:single_vs_double_LO}
\end{center}
\end{figure}

In Fig. \ref{fig:single_vs_double_LO} we
compare cross sections for the single and double-parton
scattering as a function of proton-proton center-of-mass energy. At low energies the 
conventional single-parton scattering dominates.
At low energy the $c \bar c$ or $ c \bar c c \bar c$ cross sections are much
smaller than the total cross section. At higher energies the contributions
dangerously approach the expected total cross section\footnote{New experiments at LHC will provide new input for parametrizations of the total cross section.}. This shows that inclusion of
unitarity effect and/or saturation of parton distributions may be necessary.
The effect of saturation in $c\bar c$ production has been included
but not checked versus experimental data. Presence of double-parton scattering changes the situation.
 At LHC energies the
cross section for both terms become comparable\footnote{If inclusive cross
section for $c$ or $\bar c$ was shown the cross section should be
multiplied by a factor of two -- two $c$ or two $\bar c$ in each event.}.
This is a completely new situation
when the double-parton scattering gives a huge contribution to inclusive
charm production. 

\begin{figure} [!thb]
\begin{center}
\includegraphics[width=6.5cm]{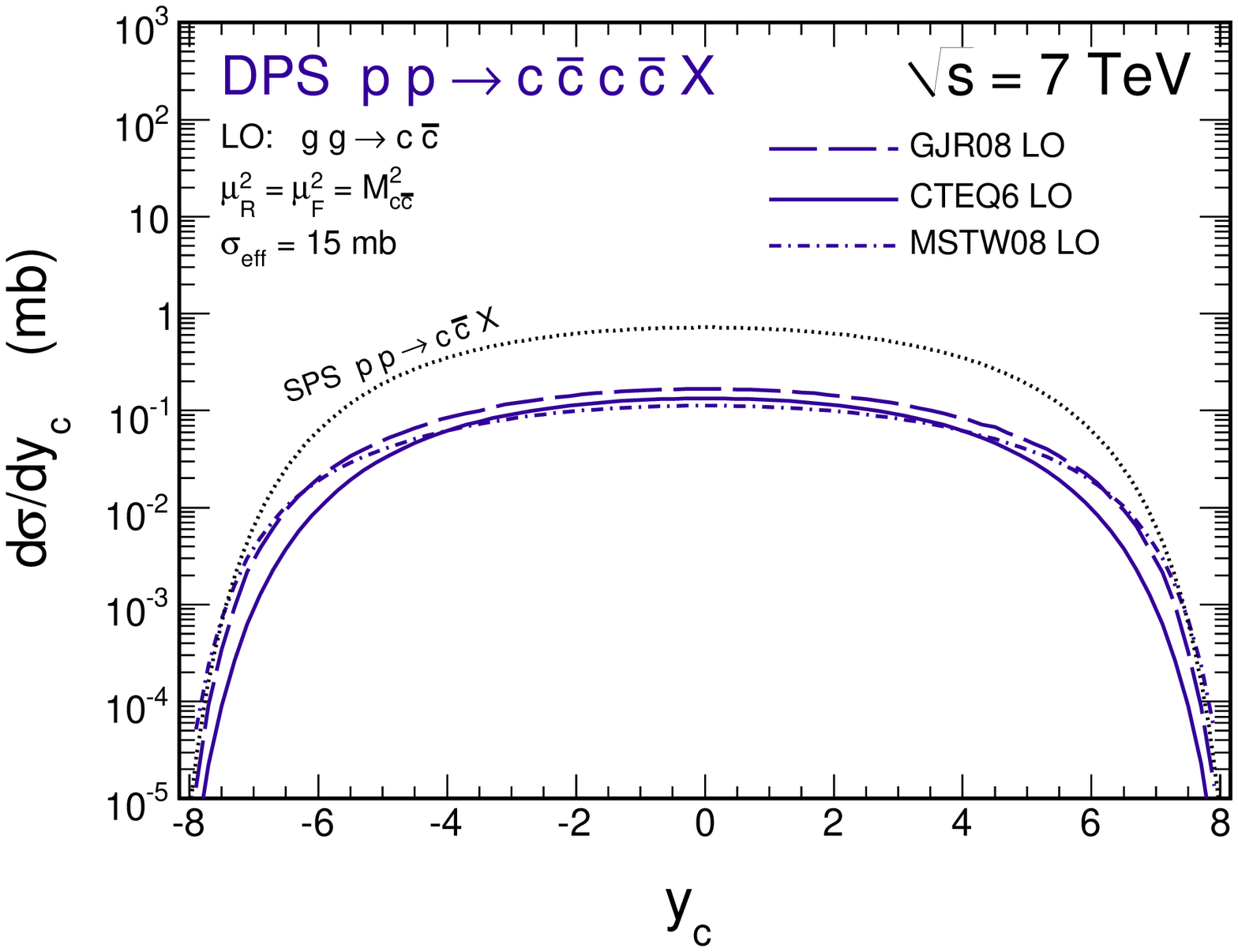}
\includegraphics[width=6.5cm]{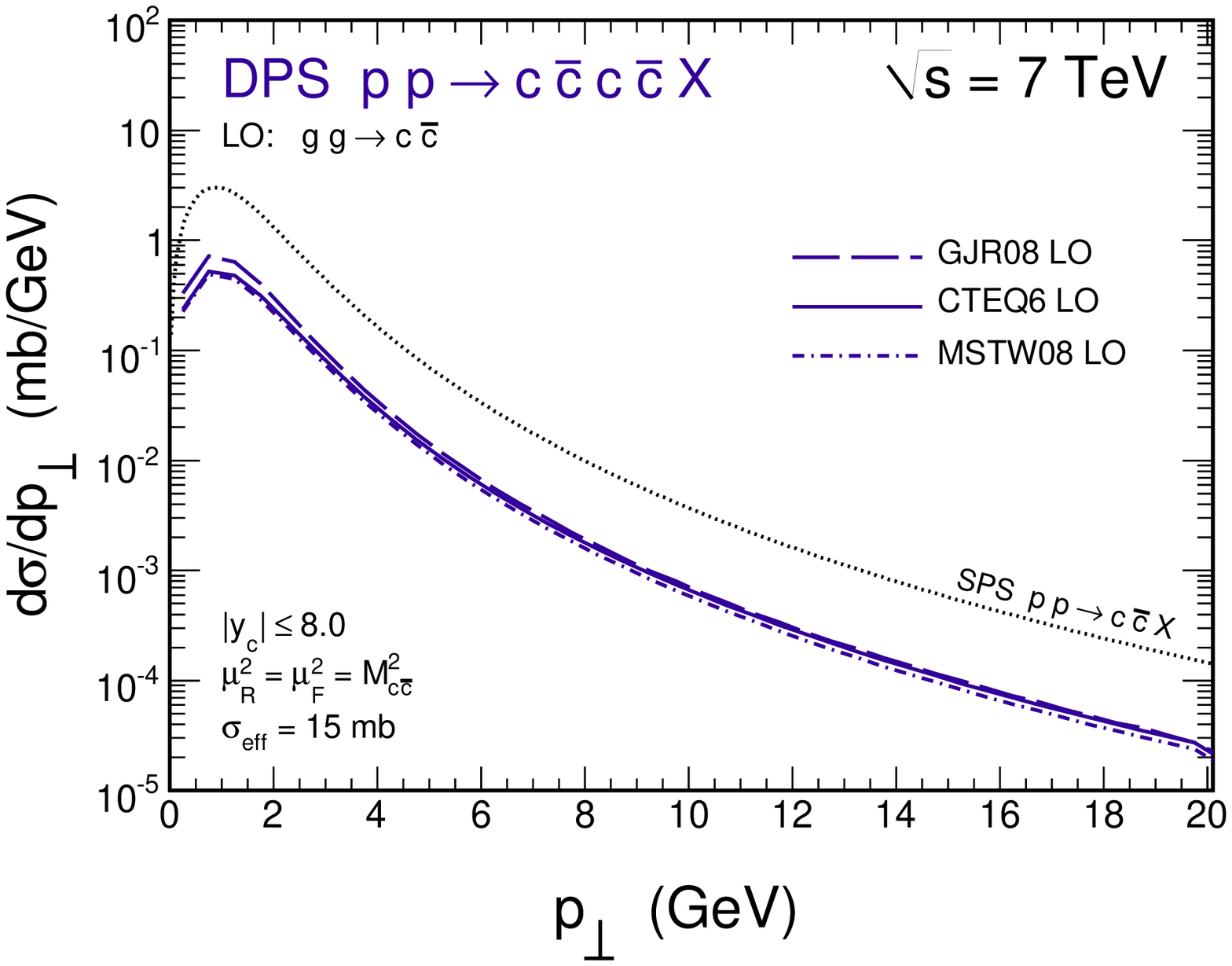}
\caption{
Distribution in rapidity (left panel) and transverse momentum (right panel) of 
$c$ or $\bar{c}$ quarks  at $\sqrt{s}$ = 7 TeV. 
}
\label{fig:double_single1}
\end{center}
\end{figure}

In Fig.~\ref{fig:double_single1},  we present
single $c$ ($\bar c$) distributions. Within approximations made in this
paper the distributions are identical in shape to single-parton
scattering distributions. This means that double-scattering contribution
produces naturally an extra center-of-mass energy dependent $K$ factor
to be contrasted with approximately energy-independent $K$-factor due to
next-to-leading order corrections. One can see a strong dependence on 
the factorization and renormalization scales which produces almost 
order-of-magnitude uncertainties and precludes a more precise estimation. 
A better estimate could be done when LHC charm data are published  and 
the theoretical distributions are somewhat adjusted to experimental data.

\end{document}